\documentclass[twocolumn]{svjour3}          % twocolumn
\smartqed  % flush right qed marks, e.g. at end of proof
\usepackage{graphicx}
\usepackage{mathptmx}      % use Times fonts if available on your TeX system
\usepackage{latexsym}
\usepackage{amsmath}
\usepackage{amssymb}
\usepackage{subcaption}
\usepackage[misc]{ifsym}
\usepackage{color}
\usepackage{epstopdf}
\usepackage{dcolumn}% Align table columns on decimal point
\usepackage{bm}% bold math
\usepackage{hyperref}% add hypertext capabilities
\hypersetup{colorlinks=true, citecolor=blue, urlcolor=blue, linkcolor=blue}
\usepackage{url}
\usepackage{threeparttable}
\begin{document}

\title{Long-term Trends of Regolith Movement on the Surface of Small Bodies}


\author{Chenyang Huang         \and
        Yang Yu                \and
        Bin Cheng              \and
        Qingyun Wang
        %etc.
}


\institute{C. Huang $\cdot$ Y. Yu $(\textrm{\Letter})$ $\cdot$ Q. Wang\at
              School of Aeronautic Science and Engineering, Beihang University, Beijing 100191, China \\
%              Tel.: +123-45-678910\\
%              Fax: +123-45-678910\\
              \email{yuyang.thu@gmail.com}
              %\email{chenyangh@buaa.edu.cn}\\
           \and
           B. Cheng \at
           School of Aerospace Engineering, Tsinghua University, Beijing 100084, China
}

\date{\today}

\maketitle

\begin{abstract}
This paper studies the long-term migration of disturbed regolith materials on the surface of Solar System small bodies from the viewpoint of nonlinear dynamics. We propose an approximation model for secular mass movement, which combines the complex topography and irregular gravitational field. Choosing asteroid 101955 Bennu as a representative, the global change of the dynamical environment is examined, which presents a division of the creeping-sliding-shedding regions for a spun-up asteroid. In the creeping region, the dynamical equation of disturbed regolith grains is established based on the assumption of ``trigger-slide'' motion mode. The equilibrium points, local manifolds and large-scale trajectories of the system are calculated to clarify the dynamical characteristics of long-term regolith movement. Generally, we find for a low spin rate, the surface regolith grains flow toward the middle latitudes from the polar/equatorial regions, which is dominated by the gradient of the geopotential. While spun up to a high rate, regolith grains tend to migrate toward the equator, which happens in parallel with a topological shift of the local equilibria at low latitudes. From a long-term perspective, we find the equilibrium points dominate the global trends of regolith movements. Using the methodology developed in this paper, we give a prospect or retrospect to the secular motion of regolith materials during the spin-up process, and the results reveal a significant regulatory role of the equilibria. Through a detailed look at the dynamical scheme under different spin rates, we achieve a macro forecast of the global trends of regolith motion during the spin-up process, which explains the global geologic evolution driven by the long-term movements of regolith materials.

\end{abstract}

\section{Introduction}
\label{sec:1}
Remote observations and in situ explorations have revealed that the natural structures of most Solar System small bodies are gravitational aggregates with unconsolidated regolith granular materials on the surface \cite{asphaug2009annual,veverka2001imaging,miyamoto2007regolith}. The regolith layers usually have a depth varying from a few meters to tens of meters and contain dispersive boulders of different sizes, which determine the intricate surface morphologies and diverse landforms of the small bodies \cite{veverka2001imaging,miyamoto2007regolith,jawin2020global}. Trails of the migrations of regolith materials have been noticed by photometric measurements in previous asteroid missions, e.g., the longitudinal ridges in the southern hemisphere of asteroid 101955 Bennu are obscured by surface materials, and large boulders indicative of toppling and downslope movement in the northern hemisphere have been identified \cite{daly2020hemispherical}. And the mass wasting on Bennu is also cogent based on the evidence that regolith material flows infill the candidate craters \cite{barnouin2019shape,walsh2019craters}. Asteroid 162173 Ryugu also exhibits unidirectional regolith deposits on imbricated flat boulders, crater wall slumping and boulder concentration in topographical lows, indicating the occurrence of mass movements \cite{sugita2019geomorphology,morota2020sample}. Likewise evidence for landslides and mass transport is observed on the surface of the Martian moon Phobos, including bright-dark streaks and lineaments of debris, etc. \cite{shi2016mass,ballouz2019surface}. These regolith layer activities highly depend on the dynamical environments near the surfaces of the asteroids, which are largely governed by the complicated geometries of the bodies and their irregular gravitational fields. Recent studies show the global patterns of the mass movements can decipher the surface evolutionary process and contribute to assessing the surface mechanical properties of small bodies \cite{cheng2021reconstructing,richardson2005global,perry2021impact}.

The migration of regolith materials on asteroids is possible to be initiated by some disturbance factors, including but not limited to global change of the surface dynamical environment due to varying spin state \cite{hirabayashi2020spin}, and exogenous excitation, such as impact events which could give rise to stress waves and seismic shaking \cite{chujo2018categorization,nishiyama2021simulation}. A slow mechanism that can modify the spin rates and obliquities of small bodies by thermal radiation forces and torques in the space environment, referred to as Yarkovsky-O'Keefe-Radzievskii-Paddack (YORP) effect \cite{bottke2006yarkovsky}, plays an important role in driving the surface evolution process of small bodies. For example, Bennu spins up with a period rate of change of $-1.02 \pm 0.15$ s per century \cite{barnouin2019shape}, inducing the centrifugal strength of surface materials slowly increasing. As a result, the dynamical slope, i.e., the supplementary angle of the angle between the outer normal vector of local surface of a small body and the geopotential force contributed by the gravitational force and centrifugal force, varies and probably exceeds the angle of repose of regolith materials. Then landslides driven by YORP spin-up occur. Another important mechanism that is suspected of driving regolith movements to modify the asteroidal surfaces is long-term seismic shaking, caused by multiple reflections and propagation of impact-induced seismic wave in the interior of bodies \cite{richardson2005global,yasui2015experimental}. For example, asteroid 433 Eros possesses a heavily cratered surface, indicating relatively frequent impact events. There exists direct evidence of mass movement such as the erasure of small craters by regolith particles \cite{veverka2001imaging}, making impact-induced seismic shaking a plausible explanation.

Different from the landslides on Earth, the movement of regolith materials induced by above disturbance factors might involve large spatial scale that is even comparable to their size \cite{jawin2020global,morota2020sample,prockter2002surface}. Low gravity level (e.g. $10^{-4}$ m$\cdot\rm{s^{-2}}$ on Ryugu and $10^{-5}$ m$\cdot\rm{s^{-2}}$ on Bennu) and low friction loss can make the intermittent regolith flow travel a long distance. Besides, the time span of mass movements is inordinately long so it is almost impossible to capture a complete event or even an ongoing process of regolith migration using on-board cameras. On the other hand, moving at a slow velocity, described as creep motion, weakens the effect of Coriolis force for regolith particles on the surface of a spinning body. A theoretical investigation of such regolith movement is confronted with the non-trivial gravitational field, complex surface topography, and an extremely large time scale. Previous studies using different methods have given impressive results, which enlighten this work and provide references. Scheeres \cite{scheeres2015landslides} explored the conditions for regolith landslides on a gravitating, rotational symmetric body using analysis method, starting from giving the basic mechanical forces acting on a stationary regolith particle. Particle movements are initiated when the local slope angle exceeds the friction angle. Gaurav \textit{et al.} \cite{gaurav2021granular} studied two-dimensional shallow granular flow on a rotating and self-gravitating ellipse applying extended avalanche dynamics, which derived the criteria of coexistence of and transformation between static and flowing regolith regimes. Cheng \textit{et al.} \cite{cheng2021reconstructing} simulated the regolith migration driven by YORP spin-up employing discrete element method and observed that creeping granular materials moved from mid-high latitudes toward low latitudes and formed a pronounced equatorial bulge. Jawin \textit{et al.} \cite{jawin2020global} documented evidence of global patterns of mass movement on the surface of Bennu based on plenty of high-resolution images taken by OSIRIS-REx cameras, unraveling the surface modification subject to regolith transport.

In this study, we propose a dynamical model that describes the long-term movement of disturbed regolith grains on the surface of a rotating asteroid, which combines the local accurate terrains and irregular gravitational field. Based on the refined modeling, global and local patterns of surface material movements are calculated and analyzed. And the nonlinear characteristics of the long-term regolith movement offer us a new perspective to understand the surface evolution process. In Section \ref{sec:2}, we describe the super slow physics of the regolith movement using a simplified model: the surface dynamical environment is formulated in analytical forms and the influences of the spin rate and friction are investigated, and then a two-stage-motion model is proposed to approximate the regolith creeping under external disturbances. In Section \ref{sec:3}, the equilibria and local manifolds of the dynamical system are calculated and analyzed. Section \ref{sec:4} presents the trajectories of surface particles at different stages of spin-up process, which exhibit the dominant roles of the equilibria on the migration of regolith materials. Section \ref{sec:5} summarizes the results of this work, and discusses their implications on secular prediction of the regolith movement at a model level, which are expected to enrich our understanding of the morphological evolution of Solar System small bodies.

\section{Modeling the movement of surface regolith particles}
\label{sec:2}
\subsection{Global construction of the surface dynamical environment}
\label{sec:2.1}
In the evolutionary process of some asteroids, variation of the centrifugal strength induced by YORP spin-up or spin-down causes global change of the surface dynamical environment, which might switch the status of regolith particles (triggering an avalanche, or stopping it). In this section, we propose a universal formulation of the planetary surface dynamical environment, and then we take asteroid 101955 Bennu as an example to show how the dynamical environment changes over the globe as the spin period of the asteroid shifts.

Considering a regolith particle on the surface of a spinning asteroid, its dynamical behavior is dominated by the gravitational force, contact force, centrifugal force, Coriolis force and other perturbation in space which are not considered in the simplified model. Based on the fact that the time scale of YORP spin-up evolution for hundred-meter asteroids is about million years \cite{barnouin2019shape}, the YORP spin-up could be treated as a quasistatic accelerating process, which means the tangential convected acceleration of a regolith particle in the body-fixed coordinate is negligible. All the physical quantities in the following formulas we listed are dimensionless. The characteristic length and time scale are adopted as the equivalent radius (the radius of a sphere with the same volume as the asteroid) and the spin period $T$ of the asteroid. The acceleration of a unit-mass particle on the surface of an asteroid is represented as Eq.\;(\ref{eq1}),
\begin{eqnarray}\label{eq1}
  \bm{a}_\mathrm{s}=\bm{g}_\mathrm{g}+\bm{f}-\bm{\omega}\times(\bm{\omega}\times\bm{r})-2\bm{\omega}\times\bm{v},
\end{eqnarray}
in which $\bm{g}_\mathrm{g}$ is the gravitation term, $\bm{f}$ is the term affected by contact force containing normal contact force and tangential friction, $\bm{\omega}$ is the vector of angular velocity of the spinning body with the magnitude $2\mathrm{\pi}$ in the dimensionless representation, and $\bm{r}$ and $\bm{v}$ respectively indicate the position vector and velocity vector in the body-fixed frame.

During the million-year evolution of asteroids, regolith materials creep down the surface at a minuscule velocity when affected by disturbance, such as impact-induced seismic activities. The creep velocity of regolith grains maintains at a small magnitude, so the Coriolis force could be negligible. The representation of the normal and tangential contact force will be given later in this Section. The gravitation and centrifugal effect could be viewed as potential force, which also determine the movement trends of stationary particles on the surface of an asteroid. The geopotential of a unit-mass particle is defined as Eq.\;(\ref{eq2}),
\begin{eqnarray}\label{eq2}
  V_\mathrm{s}=-\frac{1}{2}(\bm{\omega}\times\bm{r})\cdot(\bm{\omega}\times\bm{r})-GT^2\sigma\iiint \frac{1}{|\bm{r}-\bm{p}|}\mathrm{d}\bm{p},
\end{eqnarray}
where $G$ is the gravitational constant, $T$ is the spin period, $\sigma$ is the bulk density of the asteroid which is treated as homogenous, $\bm{p}$ is the position vector of the infinitesimal volume $\rm{d}\bm{p}$. So the gravitational and centrifugal acceleration
can be merged as the negative gradient of geopotential and Eq.(\ref{eq1}) is modified as
\begin{eqnarray}\label{eq3}
  \bm{a}_\mathrm{s}=-\nabla V_\mathrm{s}+\bm{f},
\end{eqnarray}
where $\nabla$ is the gradient operator.

The gravitational potential, i.e., the second term of Eq.\;(\ref{eq2}), and gravitation contained in the potential force in Eq.\;(\ref{eq3}) are calculated in a closed form by assuming the asteroid as a homogeneous polyhedron \cite{werner1994gravitational,werner1996exterior,wen2020hop}. Gravitation derived by polyhedron, whose calculation error only originates from the error of polyhedron shape model of an asteroid, could guarantee the accuracy whether the field point is in the exterior or interior, or on the surface of the shape model.

\begin{figure}[htbp]
  \centering
  \includegraphics[width=\linewidth]{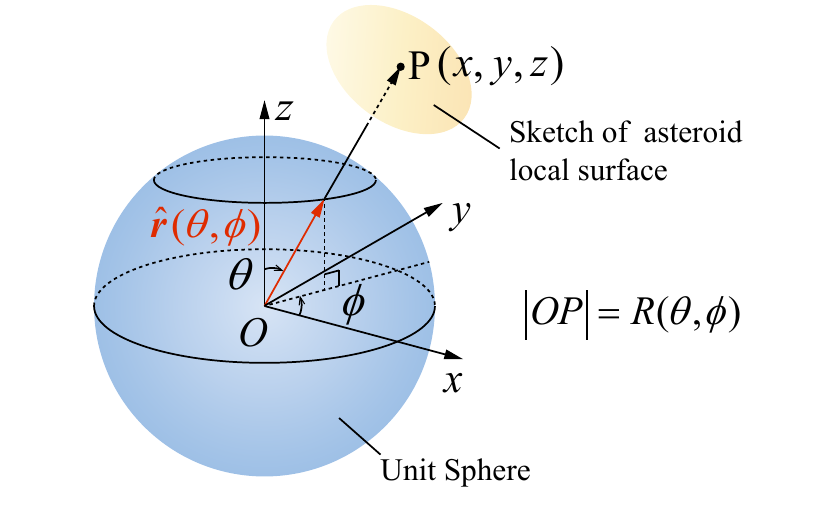}
  \caption{Cartoon showing the geometry and coordinate convention to understand the parameterized shape model of asteroids. Part of asteroid irregular surface (the light yellow surface) is drawn as example. Point $P$ on the local irregular surface can be determined uniquely by the unit orientation vector $\hat{\bm{r}}(\theta,\phi)$ and the radial distance $R(\theta,\phi)$.}
  \label{fig1}
\end{figure}

\begin{figure*}[htbp]
  \centering
  \includegraphics[width=0.8\linewidth]{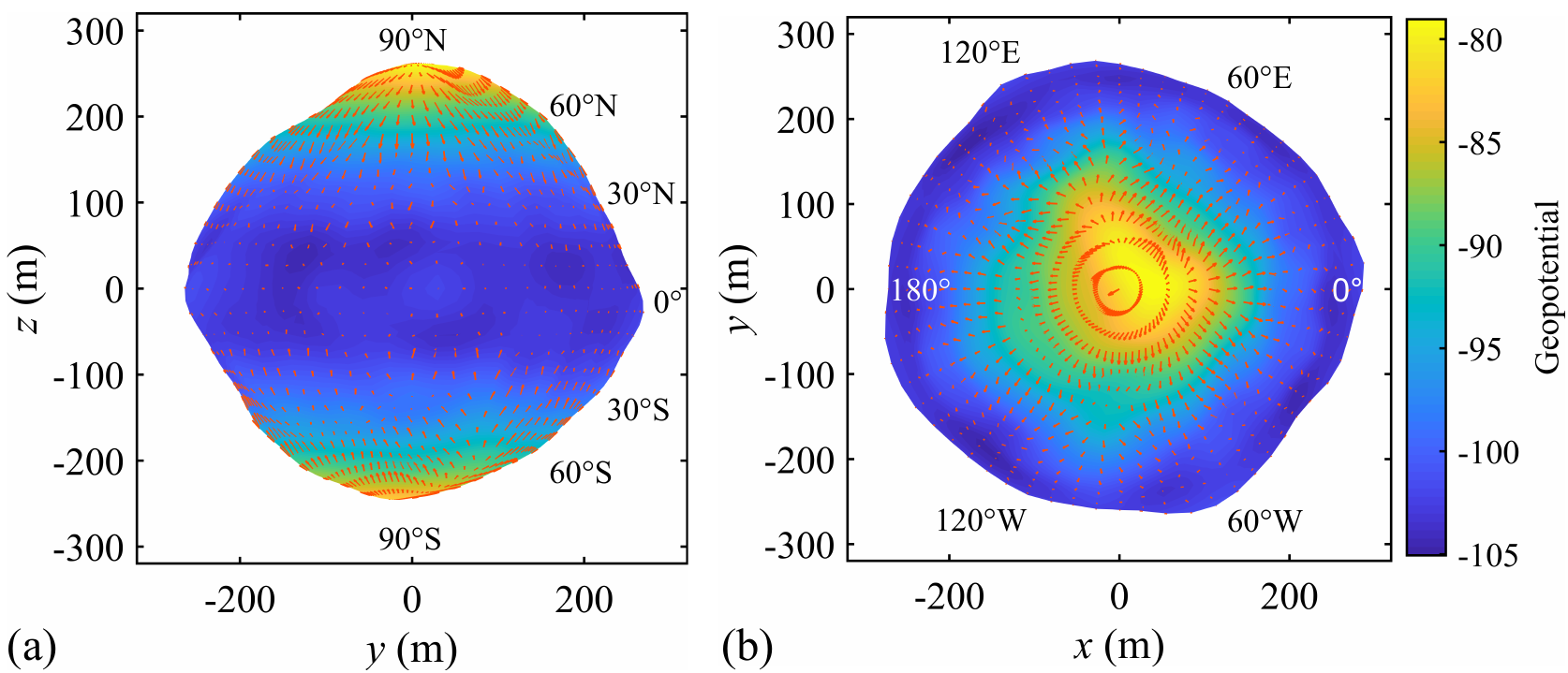}
  \caption{The shape model of Bennu in spherical harmonics. The colormap indicates the dimensionless geopotential and the arrows denote the slope directions with their length in proportion to the magnitude of $\bm{g}_\mathrm{t}$. Panel (b) is viewed from the north. Corresponding to the annotations in panel (b), the middle longitude in panel (a) is $0^{\circ}$, whose right locates east longitudes.}
  \label{fig2}
\end{figure*}

The irregular shapes of asteroids can be parameterized using spherical harmonics, which can represent the surface geometry of an arbitrary connected 3D shape with protrusions and intrusions \cite{brechbuhler1995parametrization}. In order to generate an accurate shape expression, we first establish a one-to-one mapping from a point $(x,y,z)$ on the surface of the object to its spherical coordinates $(\theta,\phi)$ and radial distance $R(\theta,\phi)$ as shown in Fig.\;\ref{fig1}. The general form of $R(\theta,\phi)$ is
\begin{eqnarray}\label{eq4}
  R(\theta, \phi)=\sum_{l=0}^{\infty} \sum_{m=-l}^{l} c_{l}^{m} Y_{l}^{m}(\theta, \phi),
\end{eqnarray}
whose expansion is based on spherical harmonic functions $Y_{l}^{m}$, representing as
\begin{eqnarray}\label{eq5}
  Y_{l}^{m}(\theta, \phi)=\sqrt{\frac{2 l+1}{4 \pi} \frac{(l-m) !}{(l+m) !}} P_{l}^{m}(\cos \theta) e^{\mathrm{i}m\phi},
\end{eqnarray}
where $P_{l}^{m}$ is the associated Legendre function. The complex coefficients $c_{l}^{m}$ of the harmonics are derived by fitting the Cartesian coordinates of the sample points on the surface of an irregular asteroid. The degree $l$ determines the spatial frequency constituents composing the parameterized shape model. In a certain range, a larger degree $l$ means that higher frequency components are included and more detailed features of the original object appear. More detailed computing methods about spherical harmonics see the appendix of \cite{yu2018dynamical}. Therefore, an arbitrary point on the surface of asteroids can be represented as
\begin{eqnarray}\label{eq6}
  \bm{R}(\theta,\phi)=R(\theta,\phi)\cdot \hat{\bm{r}}(\theta,\phi),
\end{eqnarray}
where $\hat{\bm{r}}$ is the unit orientation vector as shown in Fig.\;\ref{fig1} (Symbol $\ \hat{}\ $ indicates an unit vector throughout this paper). The tangent plane of a local surface at the position $\bm{R}(\theta,\phi)$ can be denoted by two partial derivatives of the vector function $\bm{R}(\theta,\phi)$, i.e., $\hat{\bm{R}}_\theta$ and $\hat{\bm{R}}_\phi$. The subscripts denote the partial derivative with respect to $\theta$ and $\phi$. Furthermore, the outer normal orientation of the local surface, $\hat{\bm{r}}_\mathrm{n}$, is determined by the outer product of $\hat{\bm{R}}_\theta$ and $\hat{\bm{R}}_\phi$.

The component of $-\nabla V_\mathrm{s}$ in the orientation of $\hat{\bm{r}}_\mathrm{n}$ plays a similar role to the gravity on Earth, expressed as
\begin{eqnarray}\label{eq7}
  \bm{g}_\mathrm{n}=\left( -\nabla V_\mathrm{s} \cdot \hat{\bm{r}}_\mathrm{n} \right) \hat{\bm{r}}_\mathrm{n}.
\end{eqnarray}
And the projection of $-\nabla V_\mathrm{s}$ into the tangent plane of the local surface:
\begin{eqnarray}\label{eq8}
  \bm{g}_\mathrm{t}=-\nabla V_\mathrm{s}-\left( -\nabla V_\mathrm{s} \cdot \hat{\bm{r}}_\mathrm{n} \right) \hat{\bm{r}}_\mathrm{n},
\end{eqnarray}
defines the local slope direction on account of the fact that an unconstrained particle would move along the direction of $\bm{g}_\mathrm{t}$. The contact force of a unit-mass particle on the surface refers to the plane-slider model, i.e., the normal force is equal to $\bm{g}_\mathrm{n}$ and the tangential friction is calculated using Coulomb friction model with the friction coefficient $\mu$. We divide the whole surface into several regions by first observing the orientation of vector $\bm{g}_\mathrm{n}$, and then comparing the magnitude of $g_\mathrm{t}$ and $\mu g_\mathrm{n}$. If $\bm{g}_\mathrm{n} \cdot \hat{\bm{r}}_\mathrm{n} > 0$, this location is divided into shedding region (a particle in this location will depart from the surface under the action of $\bm{g}_\mathrm{n}$); else, go to the next judging criteria: $g_\mathrm{t}-\mu g_\mathrm{n}>0,\ =0\ \textrm{and}\ <0$ correspond to the sliding region, critical boundary and creeping region respectively. The creeping region means that regolith particles therein will maintain rest if there is no disturbance, which is a region of our concern. Obviously, the shedding-sliding-creeping region division depends on the spin period $T$ and friction coefficient $\mu$.

We chose asteroid Bennu (the target object of OSIRIS-REx mission) as an example, which is a top-shaped asteroid having distinct equatorial bulge. Bennu spins at a period of about 4.30 hours with a period rate of change of $-1.02\pm0.15$ s per century \cite{barnouin2019shape}, which makes Bennu a perfect candidate for a study of the long-term trends of surface regolith migration. The shape model of Bennu in spherical harmonics is shown as Fig.\;\ref{fig2}, which is truncated to degree 20 (The sample points used for fitting the complex coefficients of the harmonics were taken from Bennu's polyhedron model \cite{nolan2013shape}). The origin is the center of mass, and the axes are the principal axes of the shape model. The $z$-axis is the spin axis. The stipulation of Bennu’s geographic coordinate is indicated by latitudes and longitudes in Fig.\;\ref{fig2}. The zero meridian and the north pole correspond to the spherical coordinate $\phi=0$ and $\theta=0$ respectively.

\begin{figure*}[htbp]
  \centering
  \includegraphics[width=0.85\linewidth]{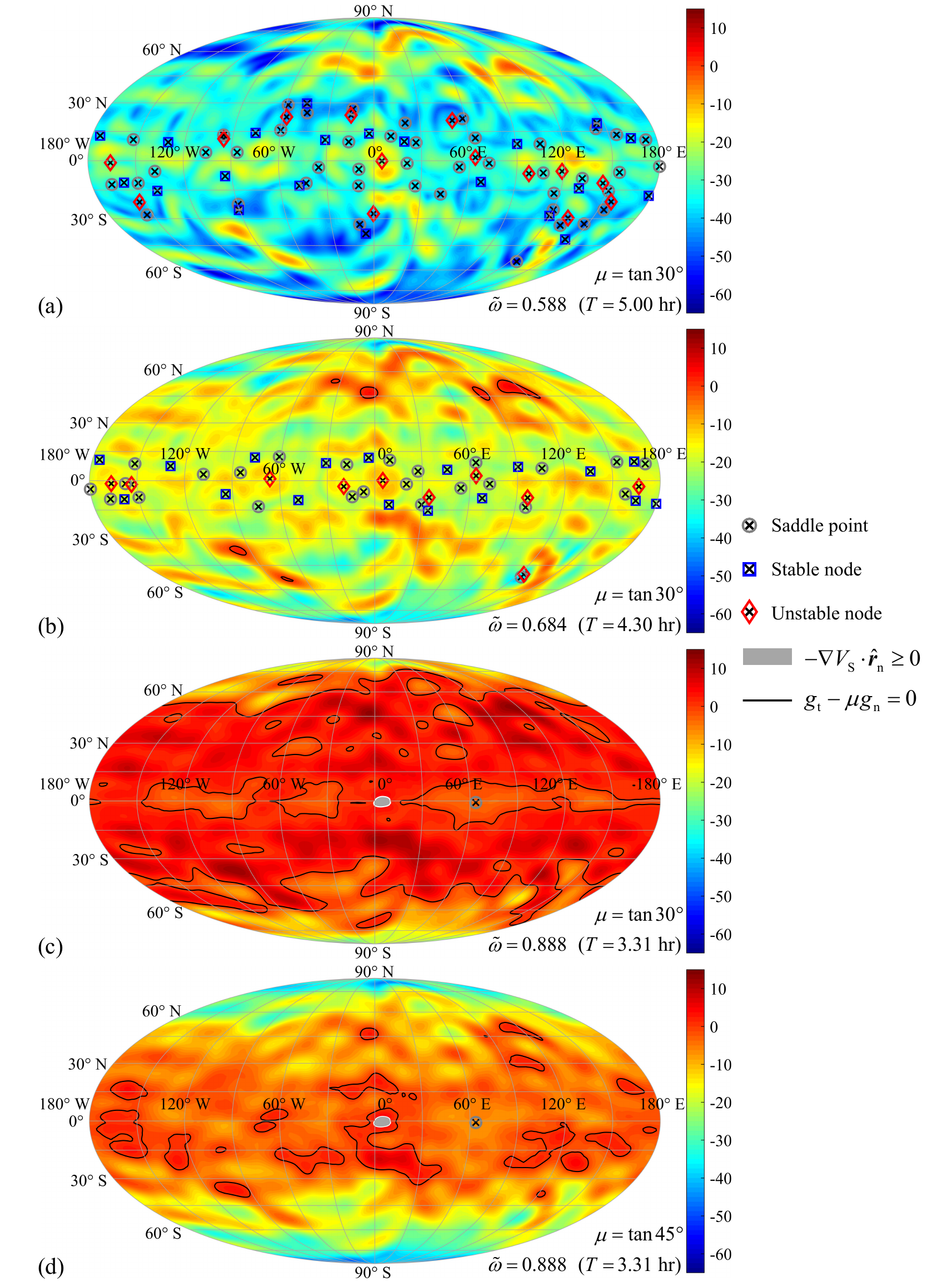}
  \caption{Global change of Bennu's surface dynamical environment. The global map is colored according to the dimensionless value of $g_\mathrm{t}-\mu g_\mathrm{n}$ and the contour $g_\mathrm{t}-\mu g_\mathrm{n}=0$ is signified in black. The gray areas in panel (c) and (d) imply the shedding region. Equilibrium points of different topological classification of the dynamical system described by Eq.\;(\ref{eq15}) are also plotted -- the circle crosses, squared crosses and rhombus crosses denote the saddle points, stable nodes and unstable nodes respectively, which will be introduced in Sec.\;\ref{sec:3}.}
  \label{fig3}
\end{figure*}

Considering a particle on the equator of a spherical asteroid with equal volume as Bennu, this particle is about to depart from the surface when the centrifugal effect weighs against the gravitation. We choose the spin rate of the spherical asteroid in this critical status as a reference value, $\omega_0=\sqrt{\frac{4}{3}\pi G \sigma}$ (corresponding to $T_0=2.94$ hr) and calculate the shedding-sliding-creeping region division based on the topography and rotational state of Bennu. Different spin rates, represented as dimensionless form $\tilde{\omega}=\omega / \omega_0$, are examined to depict the global change of dynamical region division. As shown in Fig.\;\ref{fig3}, with the increase of  $\tilde{\omega}$, the asteroid surface first changes from global creeping region to sliding-creeping coexistence, and then a shedding region emerges at the equator.  The contour of $g_\mathrm{t}-\mu g_\mathrm{n}=0$ (dark solid curves) divides the creeping region (towards the blue end of the spectrum) and the sliding region (towards the red end of the spectrum), and the shedding region is marked out in gray with white solid curves.

The atlas of Fig.\;\ref{fig3} reveals a global shifting of the dynamical environment on Bennu's surface. As the spin-up proceeding, the creeping region degenerates, which allows the regolith materials in the midlatitudes start to move spontaneously. Regolith shedding first appears at the equatorial region. Note that the sprout of shedding region at $\tilde{\omega}=0.888$, less than 1, is attributable to the conspicuous equatorial bulge of Bennu, where the radius is larger than that of the equal-volume sphere. Besides, larger friction angle obviously minishes the area of sliding region [see Fig.\;\ref{fig3}(c) and Fig.\;\ref{fig3}(d)]. In this study, we focus on the trajectory analysis of migrating grains, and the change of Bennu's shape caused by regolith movement is not considered. A study on the variation of regolith layer thickness is left to future work.

\subsection{An approximate dynamical equation of the surface particle movement}
\label{sec:2.2}
\begin{figure}[htbp]
  \centering
  \includegraphics[width=\linewidth]{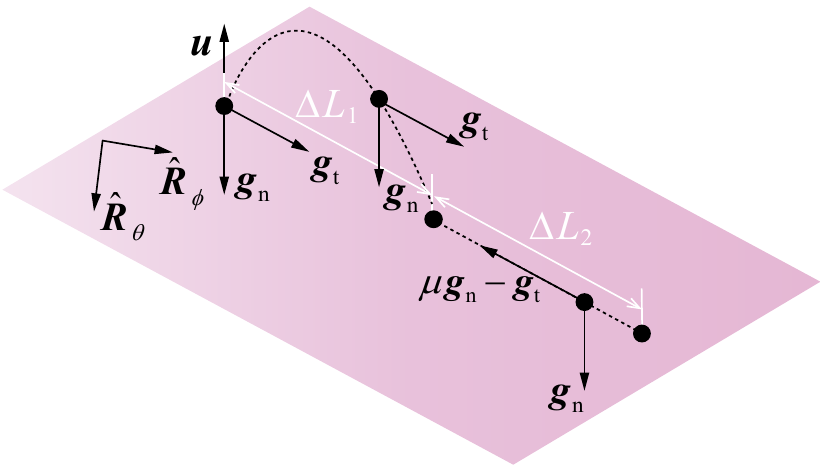}
  \caption{Approximate dynamical model of a disturbed particle. On the tangent plane of a local surface of asteroids identified by $\hat{\bm{R}}_\theta$ and $\hat{\bm{R}}_\phi$, the ``trigger-slide'' process of a disturbed regolith particle is simplified as two stages: the projectile motion and decelerating motion.}
  \label{fig4}
\end{figure}

Regolith particles in the creeping region ($g_\mathrm{t}-\mu g_\mathrm{n}<0$) might start to move subject to the disturbance in the environment, such as long-term geological activities induced by impact events. Different from the slope angle [its value is equal to $\mathrm{arctan}(g_\mathrm{t}/g_\mathrm{n})$] variation induced by YORP spin-up, small perturbations like stress waves or seismic shaking could make regolith materials leave from the surface temporarily and slightly. Hence, we propose a dynamical approximation to this ``trigger-slide'' process, a two-stage model covering projectile motion left the surface and decelerating motion on the surface as shown in Fig.\;\ref{fig4}. The main body of the asteroid is viewed as a rigid body covered with regolith materials, and their interface spontaneously exerts disturbances on the regolith grains, causing them gravitated toward the potential lows like sand in the Chladni plate experiment \cite{kopitca2021programmable}. We first establish the equation of a disturbed particle in the creeping region.

When a particle obtains an initial velocity to leave the surface, which is assumed to be perpendicular to the local surface, it will fall down affected by $\bm{g}_\mathrm{n}$ and go ahead under the action of $\bm{g}_\mathrm{t}$ (see Fig.\;\ref{fig4}). Once the particle has landed again, it will be influenced continuously by friction until its movement ceases, causing a deceleration $\mu \bm{g}_\mathrm{n}-\bm{g}_\mathrm{t}$. Representing the forward distance in the slope direction of these two stages as $\mathrm{\Delta} L_1$ and $\mathrm{\Delta} L_2$, they can be calculated as
\begin{eqnarray}\label{eq9}
  \mathrm{\Delta} L_1 = \frac{1}{2} g_\mathrm{t} \mathrm{\Delta} t_1^2,
  \mathrm{\Delta} L_2 = \frac{\left( g_\mathrm{t} \mathrm{\Delta} t_1 \right) ^2}{2\left( \mu g_\mathrm{n}-g_\mathrm{t} \right)},
\end{eqnarray}
in which $\mathrm{\Delta} t_1$ and $\mathrm{\Delta} t_2$ correspond to the time in two stages. And $\mathrm{\Delta} t_2$ is proportional to $\mathrm{\Delta} t_1$:
\begin{eqnarray}\label{eq10}
  \mathrm{\Delta} t_2 = \frac{g_\mathrm{t}\mathrm{\Delta} t_1}{\mu g_\mathrm{n}-g_\mathrm{t}}.
\end{eqnarray}
Introducing $\mathrm{\Delta} \tau = \left( \mathrm{\Delta} t_1+\mathrm{\Delta} t_2 \right) ^2$ to characterize the scaled time, define the normalized velocity as the ratio of the distance to the scaled time:
\begin{eqnarray}\label{eq11}
  v = \underset{\mathrm{\Delta} \tau \rightarrow 0}{\lim}\frac{\mathrm{\Delta} L}{\mathrm{\Delta} \tau}=\frac{g_\mathrm{t}\left( \mu g_\mathrm{n}-g_\mathrm{t} \right)}{2\mu g_\mathrm{n}},
\end{eqnarray}
where $\mathrm{\Delta} L=\mathrm{\Delta} L_1+\mathrm{\Delta} L_2\nonumber$. We propose a vector function $\bm{v}$ to measure the velocity of disturbed particles, whose magnitude is $v$ and orientation is along the slope direction $\bm{g}_\mathrm{t}$ which determines the steepest descent direction of the geopotential on the surface. $\bm{v}$ could also be derived in terms of the derivative of position coordinate with respect to the scaled time $\tau$, i.e.,
\begin{eqnarray}\label{eq12}
  \bm{v}=v \hat{\bm{g}_\mathrm{t}}=\frac{\mathrm{d}\bm{R}\left( \theta ,\phi \right)}{\mathrm{d}\tau},
\end{eqnarray}
where $\hat{\bm{g}_\mathrm{t}}$ is the unit vector in the slope direction. Project $\bm{v}$ onto the direction of $\hat{\bm{R}}_{\theta}$ and $\hat{\bm{R}}_{\phi}$:
\begin{eqnarray}\label{eq13}
  v \hat{\bm{g}_\mathrm{t}} = \frac{g_\mathrm{t}\left( \mu g_\mathrm{n}-g_\mathrm{t} \right)}{2\mu g_\mathrm{n}}\left[ (\hat{\bm{g}}_\mathrm{t} \cdot \hat{\bm{R}}_{\theta}) \hat{\bm{R}}_{\theta}+(\hat{\bm{g}}_\mathrm{t} \cdot \hat{\bm{R}}_{\phi}) \hat{\bm{R}}_{\phi} \right],
\end{eqnarray}
\begin{eqnarray}\label{eq14}
  \bm{R}' = \bm{R}_{\theta} \theta'+\bm{R}_{\phi} \phi' = |\bm{R}_{\theta}|\theta' \hat{\bm{R}}_{\theta}+|\bm{R}_{\phi}|\phi' \hat{\bm{R}}_{\phi}.
\end{eqnarray}
The operator $'$ in this paper represents ${\mathrm{d}}/{\mathrm{d}\tau}$. Contrasting the magnitude in the direction of $\hat{\bm{R}}_{\theta}$ and $\hat{\bm{R}}_{\phi}$ of these two representations, the first-order differential equations about $\theta$ and $\phi$ are acquired and expressed as
\begin{equation}\label{eq15}
  \left[ \begin{array}{c}
	f\left( \theta ,\phi ;\mu ,\tilde{\omega} \right)\\
	g\left( \theta ,\phi ;\mu ,\tilde{\omega} \right)\\
  \end{array} \right] :=
  \left[ \begin{array}{c}
	\theta'\\
	\phi'\\
  \end{array} \right] =
  \frac{\left( \mu g_\mathrm{n}-g_\mathrm{t} \right)}{2\mu g_\mathrm{n}}
  \left[ \begin{array}{c}
	\bm{g}_\mathrm{t}\cdot \bm{R}_{\theta}/|\bm{R}_{\theta}|^2\\
	\bm{g}_\mathrm{t}\cdot \bm{R}_{\phi}/|\bm{R}_{\phi}|^2\\
  \end{array} \right].
\end{equation}

Equation\;(\ref{eq15}) governs the movement of a grain driven by surface triggering events and determines a global flow map on the shape model of an asteroid. Considering the spin-up time scale of an asteroid, the whole migrating process of a regolith grain usually takes a long term, thus the flow map can be regarded as a prediction of the secular trends of regolith movement. In the model as described by Eq\;(\ref{eq15}), the friction $\mu$ and spin rate $\tilde{\omega}$ are two major parameters that influence the dynamical behaviours of the whole system. The rest part of this paper will take a detailed look at them.

\section{The equilibria and local manifolds}
\label{sec:3}
The equilibrium points and the local manifolds in their vicinity reflect the dynamical characteristics of the system. The equilibria of Eq\;(\ref{eq15}) are obtained by solving
\begin{eqnarray}\label{eq16}
  \left[ \begin{array}{c}
	f\left( \theta ,\phi ;\mu ,\tilde{\omega} \right)\\
	g\left( \theta ,\phi ;\mu ,\tilde{\omega} \right)\\
  \end{array} \right] = \bm{0}.
\end{eqnarray}
While searching for the exact solution of Eq\;(\ref{eq16}), we find the zeros of $f$, $g$ essentially depend on two simpler equations, i.e., $\mu g_\mathrm{n}-g_\mathrm{t}=0$ and $g_\mathrm{t}=0$. As shown in Fig.\;\ref{eq3}, $\mu g_\mathrm{n}-g_\mathrm{t}=0$ depicts the boundaries of creeping region, which are continuous curves. And the isolated equilibria are determined by equation $g_\mathrm{t}=0$, which locate in the creeping region and dominate the migrating trajectories. As a numeric scheme, we first calculate the values of $g_\mathrm{t}$ on the nodes of a global grid $(\theta_i,\phi_j)\ (i,j=1,2,...;\ 0<\theta_i<\mathrm{\pi},\ 0<\phi_j<2\mathrm{\pi})$. Then we screen out the nodes $(\theta_i,\phi_j)$ on which $g_\mathrm{t}$ is sufficiently small as the initial guesses for the next iteration. Note that numerically not all equilibrium points are guaranteed to be found. We use dense grid and relaxed threshold values to improve the coverage of our scheme. Through repeated attempts of grid density and iteration settings, we located the equilibrium points over the globe for different system parameters, $\mu$ and $\tilde{\omega}$. These zero-velocity points are plotted using black ``$\times$" in Fig.\;\ref{fig3}, Fig.\;\ref{fig7} and Fig.\;\ref{fig8}. The circles, squares and rhombus covering the black ``$\times$" denote the topological classification of the equilibrium points (see next paragraph for a detailed discussion).

The stability of the equilibrium points is examined pursuant to the linear approximate system of Eq.\;(\ref{eq15}), i.e.,
\begin{eqnarray}\label{eq17}
  \left[ \begin{array}{c}
	\theta'\\
	\phi'\\
  \end{array} \right] =
  \left[ \begin{array}{cc}
    \frac{\partial f(\theta^*,\phi^*)}{\partial \theta} & \frac{\partial f(\theta^*,\phi^*)}{\partial \phi} \\
    \frac{\partial g(\theta^*,\phi^*)}{\partial \theta} & \frac{\partial g(\theta^*,\phi^*)}{\partial \phi}
  \end{array} \right]
  \left[ \begin{array}{c}
  \theta \\
  \phi \\
  \end{array} \right],
\end{eqnarray}
in which small quantities of the expansion form above the second order are negligible. The system properties are approximated using the linearized coefficient matrix \cite{leine2010historical,ghaffari2015new}, in which the superscript $^*$ denotes the coordinates of the equilibrium point. The eigenvalues and eigenvectors of the linearized coefficient matrix in Eq.\;(\ref{eq17}) are solved numerically. The local topology of each equilibrium point can be determined from the spectrum of the eigenvalues, which leads to a classification of the equilibrium points \cite{jiang2016order}. In all the cases considered in this study, i.e., Bennu's shape with varying $\tilde{\omega}$ and $\mu$, we find only three topological types of equilibria, the saddles (marked in circled crosses), stable nodes (marked in squared crosses) and unstable nodes (marked in rhombus crosses), as shown in Fig.\;\ref{fig3}, Fig.\;\ref{fig7} and Fig.\;\ref{fig8}. The topological equivalence between the linearized system and the original system is guaranteed by theory for these equilibrium types, which enables us to understand the local behaviours of the creeping flow around the equilibria from the eigen-structure of Eq.\;(\ref{eq17}). Table\;\ref{tab1} lists the locations of three representative equilibria, together with the geographic coordinates and the corresponding eigenvalues of the coefficient matrix, calculated using $\mu=\mathrm{tan}30^{\circ}$ and $\tilde{\omega}=0.684$ ($T=4.30$ hr). Equilibria $E_1$, $E_2$ and $E_3$ all possess two real eigenvalues: $E_1$ has one positive and one negative, determining a saddle point; $E_2$ has two negative, determining a stable node; $E_3$ has two positive, determining an unstable node. Among the three types, only the stable node $E_2$ is asymptotically stable (all negative real eigenvalues), and the other two types are unstable, which applies to the original nonlinear system.

\begin{table}[htbp]
  \caption{Locations of three representative equilibria and eigenvalues of the linearized coefficient matrix}
  \label{tab1}
  \begin{tabular}{cllll}
    \hline\noalign{\smallskip}
    Equil. Pt.  &($\theta$, $\phi$) & Geo. Coord. & $\lambda_1$ & $\lambda_2$ \\
    \noalign{\smallskip}\hline\noalign{\smallskip}
    $E_1$ & (1.74, 3.35) & (9.41$^\circ$S, 168.06$^\circ$W) & -36.30 & 25.63 \\
    $E_2$ & (1.37, 6.21) & (11.75$^\circ$N, 3.94$^\circ$W) & -83.44 & -24.42\\
    $E_3$ & (2.45, 2.17) & (50.23$^\circ$S, 124.10$^\circ$E) & 24.27 & 115.17\\
    \noalign{\smallskip}\hline
    \end{tabular}
    \begin{tablenotes}
        \footnotesize
        \item * Equil. Pt. and Geo. Coord. in the first line are the abbreviations for Equilibrium Point and Geographic Coordinate respectively.
    \end{tablenotes}
\end{table}

\begin{figure}[htbp]
  \centering
  \includegraphics[width=0.75\linewidth]{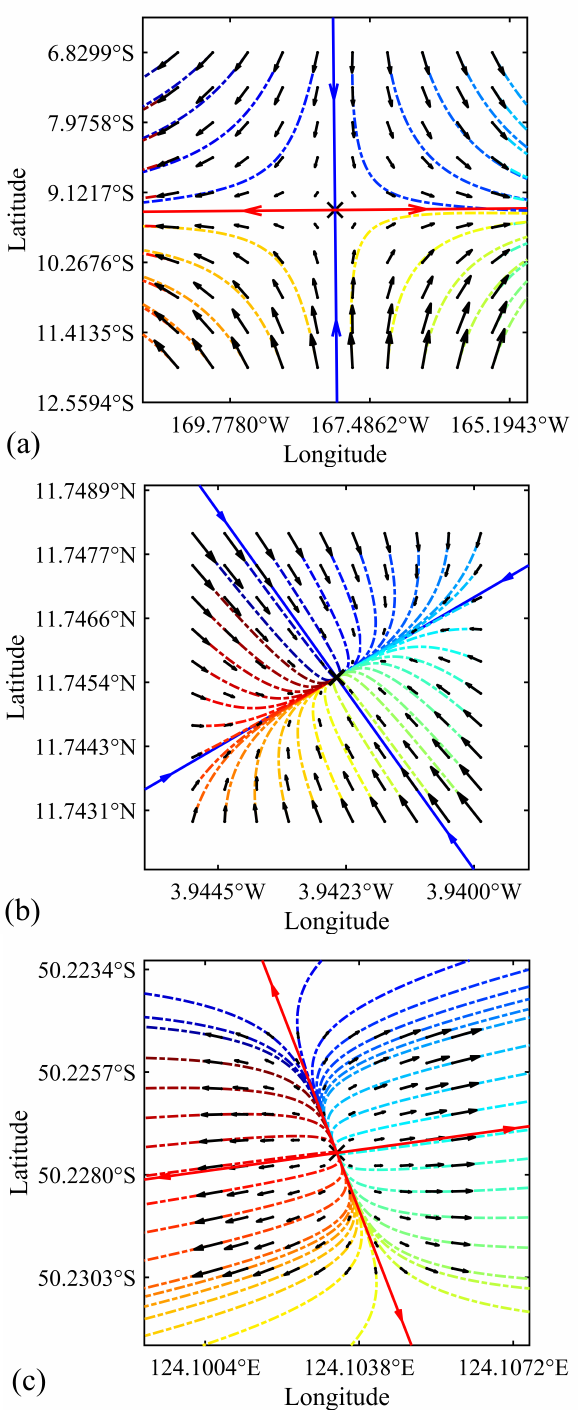}
  \caption{Local manifolds of equilibria $E_1$, $E_2$ and $E_3$. Black markers ``$\times$" in three panels locate $E_1$, $E_2$ and $E_3$ respectively. Invariant subspace of the linearized system is plotted, blue lines and arrows portraying stable subspace and the red corresponding to unstable subspace. Local vector fields and trajectories of the original nonlinear system are depicted using black arrows (whose length is proportional to the magnitude of local normalized velocity) and rainbow colored dash-dot lines respectively.}
  \label{fig5}
\end{figure}

Figure\;\ref{fig5} spotlights the local manifolds around the equilibria, which determine the local behaviours of the triggered regolith grains. The stable invariant subspace of the linearized system is spanned by the eigenvectors corresponding to negative eigenvalues, and the unstable subspace, by the eigenvectors corresponding to positive eigenvalues \cite{jiang2015topological}. In the local zoomed Fig.\;\ref{fig5}, the stable manifolds (blue solid lines) and the unstable manifolds (red solid lines) of equilibria $E_1$, $E_2$ and $E_3$ are illustrated using the eigenvectors of the linearized system, and the flow vector fields (black arrows) and migration paths (dash-dot curves in rainbow color) are calculated in terms of the original nonlinear system. The vicinal flow characteristics of equilibria of different topological classifications are obviously distinct. For the saddle point, the trajectories approach the equilibrium along the direction of the vector of stable subspace and move away from it following the direction of the vector of unstable subspace, implying a regulatory effect of the saddle point on the nearby regolith migration. For the stable and unstable nodes, the trajectories consistently approach and depart from the equilibria, respectively, which indicate the convergent and emanative movement tendencies of the surface regolith materials in the vicinity.

\begin{figure*}[htbp]
  \centering
  \includegraphics[width=\linewidth]{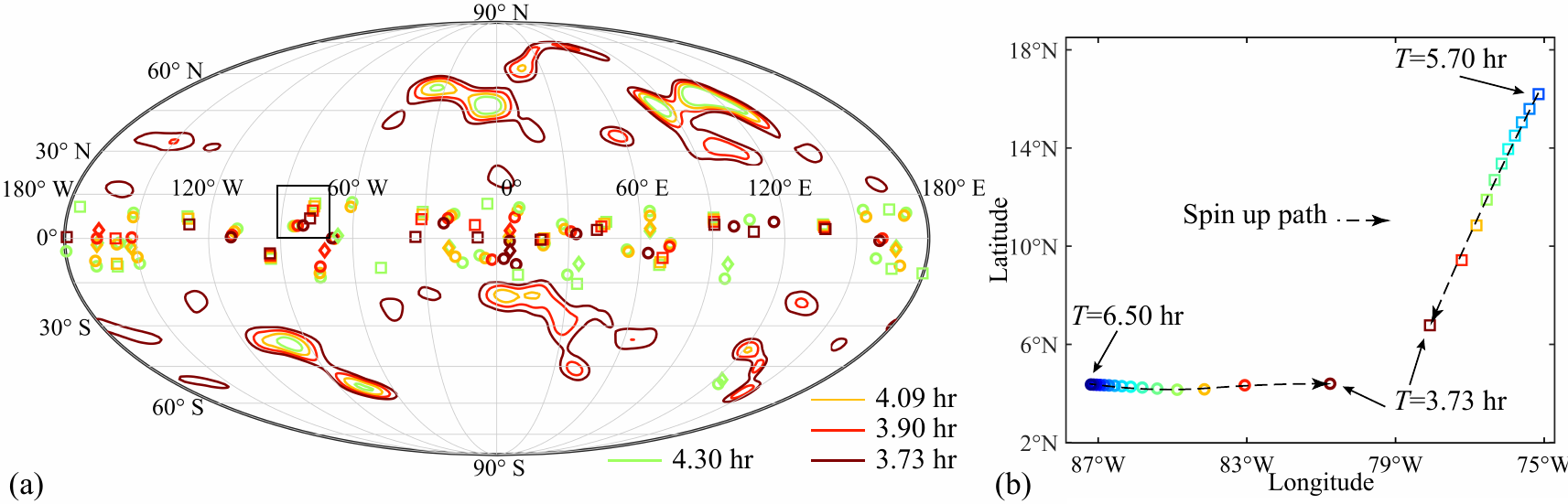}
  \caption{Equilibrium evolution during spin-up process. (a) Dividing lines between the sliding and creeping regions and equilibria evolve along the spin-up path ($\tilde{\omega}$=0.684, 0.719, 0.754 and 0.788, i.e., $T$=4.30, 4.09, 3.90 and 3.73 hr), distinguished by four colors. Circles, squares and rhombuses respectively represent saddle points, stable nodes and unstable nodes. (b) Full-process evolution of two equilibrium point chains. The spin-up path is extended to present the evolution of equilibrium points in the black box of panel (a), from their generation to annihilation.}
  \label{fig6}
\end{figure*}

We examine the location and topology shifting of equilibria on the surface as the asteroid spins up under YORP effect. In a spin-up path from $\tilde{\omega}=0.684$ ($T=4.30$ hr) to $\tilde{\omega}=0.788$ ($T=3.73$ hr), the evolution of equilibrium points is displayed using the results of four different spin rates, which are marked in different colors as shown in Fig.\;\ref{fig6}(a). The curves show the contour of $g_\mathrm{t}-\mu g_\mathrm{n}=0$, i.e., the boundary between the creeping and sliding region. We observe continuous changes in the location of equilibrium points, which look like chains.  For instance, two equilibrium point chains are formed in the black box of Fig.\;\ref{fig6}(a) as the equilibrium points constantly shift during the spin-up process. Figure\;\ref{fig6}(b) calculated this evolutionary path in greater detail: the saddle point and stable node first generate at $T=6.50$ hr and $T=5.70$ hr, respectively, and both annihilate at $T=3.73$ hr when the two chains nearly collide. The enlarged view panel (b) demonstrates the full-process migration of equilibria in the extended spin-up path. Besides, we notice some equilibrium chains gradually vanish in the $\tilde{\omega}$ range we checked, and some emerge from vain, which exhibit the complex interaction between the gravitational potential and centrifugal potential on the asteroid's irregular surface. However, a common tendency can still be found: as the asteroid spun up, the total number of equilibrium points decreases and their locations become more concentrated around the equator. Because for high spin rates the small-scale topographic relief has little contribution to the geopotential gradient, which is dominated by the overall top-like shape of Bennu.

\section{Global migration tendency under the regulation of the equilibria}
\label{sec:4}
\begin{figure}[htbp]
  \centering
  \includegraphics[width=\linewidth]{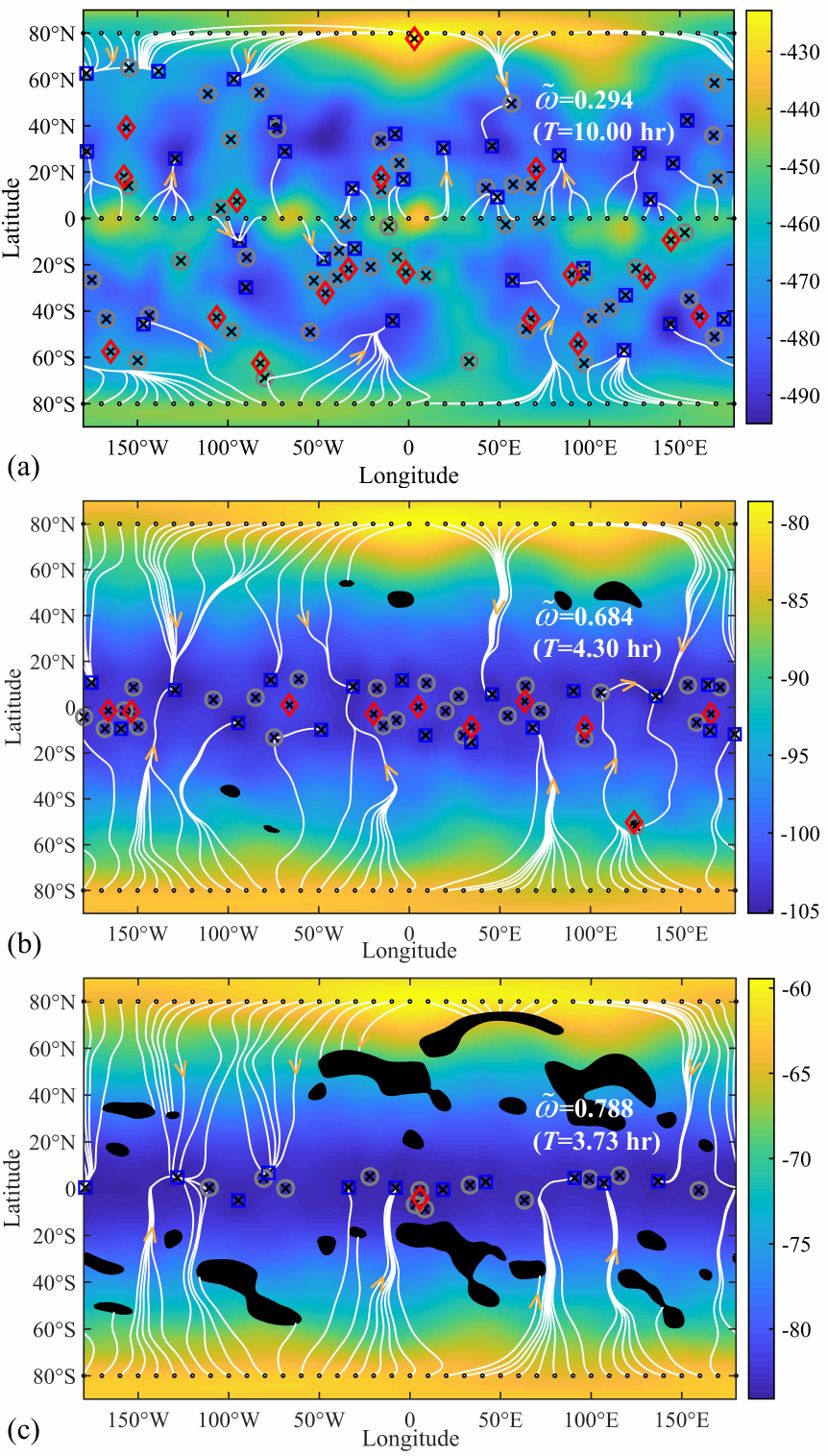}
  \caption{Trajectories of high-latitude and equatorial releasing regolith particles. Releasing positions are marked using small black circles and the white lines denote the trajectories of disturbed particles. Orange arrows are set to assist in identifying the migrating direction. The colormap indicates the dimensionless geopotential of Bennu at different spin period. And the sliding regions are distinguished in black. The friction angle in these cases is $30^{\circ}$. The equilibrium point markers represent the same topological categories as those in Fig.\;\ref{fig3}.}
  \label{fig7}
\end{figure}

Solutions of Eq.\;(\ref{eq15}) are obtained numerically, which outline the large-scale trajectories of the regolith particles in the creeping region. Considering a quasistatic migrating process of the regolith materials (see Sec.\;\ref{sec:2}), we realize the time span for a whole migrating trajectory could be extremely long, and these trajectories can roughly be regarded as a prediction of the long-term tendency of the regolith migration. We notice that the global behaviors of these trajectories are obviously regulated by the equilibria, whose locations and topologies are changing as the asteroid is spun up. The migrating trajectories under different spin rates are then calculated. When the spin rate is slow, e.g., $\tilde{\omega}=0.294$ ($T=10$ hr), high geopotential locates at high latitudes and in the equatorial regions [see Fig.\;\ref{fig7}(a)]. So when some locations at high latitudes and in the equatorial regions are set to the initial values of trajectory integration (called ``releasing positions" in the following), particle movements toward mid-latitudes are observed. At current spin period 4.30 hr, disturbed high-latitude materials move toward equatorial regions as shown in Fig.\;\ref{fig7}(b). Except for flowing toward the equator, some trajectories stop at the boundary of sliding regions at higher spin rate [see Fig.\;\ref{fig7}(c)], which means that Eq.\;(\ref{eq15}) is no longer active and surface materials could spontaneously move because of $g_\mathrm{t}>\mu g_\mathrm{n}$. In consistent with the local dynamical characteristics near the equilibria plotted in Fig.\;\ref{fig5}, global trajectories released from high latitudes and equatorial regions change their direction when encountering the saddle points (approach and then depart from the saddle points) and converge in the vicinity of the stable nodes. Besides, we remark that the friction coefficient $\mu$ have no influence on the locations of equilibria, because the parameter $\mu$ only appears in the scalar coefficient term of Eq.\;(\ref{eq15}). And calculations show that different $\mu$ will not change the topological classification of equilibria. This fact leads to a deduction that the regolith movement in the creeping region is independent on the friction between the regolith grains, which favours a common evolutionary path for asteroids of different chemical compositions.

Next we attempt to interpret the change of regolith movement trends during spin-up process. As the spin rate increases, the creeping region is gradually minishing, resulting in the variation of accessible area for the moving regolith materials overall. When the asteroid spins at an extremely low angular rate, the gravitational potential is dominated in the geopotential. In topographical highs (where the elevation is high), such as the equatorial bulges, polar regions and local highlands, the gravitational potential is at a relative high level. Hence, the colormap of the geopotential displays patchy distributions, such as Fig.\;\ref{fig7}(a). In accordance with the gradient of the geopotential, the movements of regolith particles from polar regions and equatorial regions toward middle latitudes are observed in the calculating results, coexisting with localized flow induced by local terrain. Therefore, zeros of normalized velocity, i.e., the equilibrium points, distribute in a large latitude range and are of large quantity, as shown in Fig.\;\ref{fig7}(a). Observations have proved signs of mass movement from the equatorial ridge toward the higher latitudes on the surface of Ryugu (a top-shaped asteroid, with an equatorial radius of 502 m and a period of 7.63 hr \cite{watanabe2019hayabusa2}), which are consistent with the current geopotential of Ryugu \cite{sugita2019geomorphology}. When the spin rate reaches a high level, the effect of the centrifugal potential in the geopotential is powerful, which even makes the geopotential level parallel with the latitude and decreases the geopotential difference at the same latitude. As a result, the global flow trends are uniformly toward the equator and the equilibrium points are lessened and concentrate on the equator, as shown in Fig.\;\ref{fig7}(c).

\begin{figure*}
  \centering
  \includegraphics[width=0.8\linewidth]{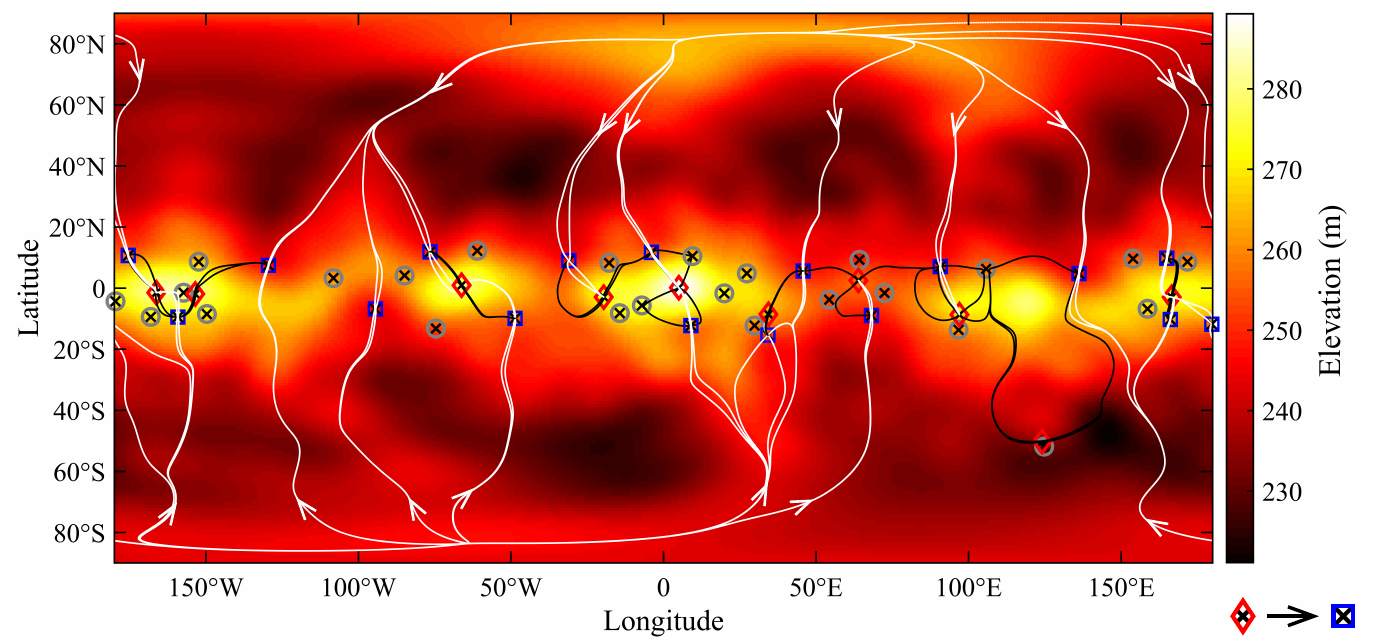}
  \caption{Trajectories flowing toward stable nodes (white lines) and departing from unstable nodes (black lines) at current spin period 4.30 hr and $\mu=\mathrm{tan}30^{\circ}$. White arrows indicate the trajectory direction. And the trajectories between stable and unstable nodes are uniformly from the unstable to the stable, as shown in the schematic at the bottom right. The colormap indicates the elevation distribution of Bennu. The equilibrium point markers represent the same topological categories as those in Fig.\;\ref{fig3}.}
  \label{fig8}
\end{figure*}

Special attention is paid to the trajectories connected to the vicinity of the equilibria, i.e., as the scaled time $\tau$ increases, the migrating regolith particles are approaching or departing from the equilibria under the regulation of the stable/unstable manifolds. For example, we check the trajectories stretching out from the vicinity of stable nodes with descending integral time series at spin period $T=4.30$ hr (the observed rotational period of Bennu), suggesting the historical migrating paths of regolith materials toward the stable nodes.  As shown by the white lines in Fig.\;\ref{fig8}, convergent migrating trajectories of regolith materials around the stable nodes in the equatorial region are mainly from high-latitude regions and the vicinity of the unstable nodes at low latitudes. The origins at high latitudes are proved to be unstable nodes in terms of the supplementary verification. Similarly, the trajectory integrations initialized from the vicinity of unstable nodes with respect to ascending $\tau$ are also calculated to track the movement paths of regolith materials. Uniformly, emanative materials flow toward nearby stable nodes along the black lines. Not only in Fig.\;\ref{fig8} but also in Fig.\;\ref{fig7}, saddle points have certain effects on the trajectories, such as deflecting flow direction. Generally speaking, the global mass movement reveals an overall tendency toward equators, coexisting with the redistribution of regolith materials at low latitudes. The calculating results for Bennu at current spin period have been supported by the identifications of mass movements based on the mission images \cite{jawin2020global}. Combined with the elevation colormap in Fig.\;\ref{fig8}, in low-latitude region stable nodes locate at relatively low-lying areas while unstable nodes are in topographic highs. In some degree, the regolith migration from unstable nodes to stable nodes might reduce the difference in terrain height.

From a long-term perspective, the equilibrium points dominate the global trends of regolith movements. Using the methodology developed in this work, we can give a prospect or retrospect to the secular motion of regolith materials, by tracking the solution trajectories forward or backward along the time axis. The three types of equilibria prove to play a significant regulatory role: the migrating regolith grains converge and end up in the vicinity of the stable nodes, and escape from the unstable nodes; for saddle points, the situation is more complex because the trajectories in different directions show opposite migration trends due to their distance from the stable manifold or the unstable manifold respectively. In a typical case, when a trajectory is approaching the saddle along the stable manifold, it will be deflected rapidly as it passing by, exhibiting a ``slingshot'' motion. Combining the discussion on the evolution of equilibria in Sec.\;\ref{sec:3}, we achieve a macro forecast of the global trends of regolith motion during the spin-up process, i.e., the generation, migration and annihilation of the equilibria imply the turning points of global geologic processes, which are driven by the long-term migration of surface regolith materials.

\section{Conclusion}
\label{sec:5}
In this paper, we establish a dynamical model to describe the long-term migration of regolith materials on the surface of small bodies, considering the influences from the complex shape and irregular gravitational field of the small body. We analyze the global change of the surface dynamical environment during the spin-up process and for different surface friction coefficient. The approximate dynamical equation of disturbed regolith grains is obtained by assuming a slowly intermittent migrating pattern, which governs the long-term movement of surface regolith materials on a specific asteroid. Taking asteroid Bennu as an example, we examine the equilibrium points of the creeping motion, and the large-scale migrating trajectories of the regolith materials. The local manifolds of the equilibria and the morphology of the large-scale trajectories reveal a detailed map that manifests the long-term geological evolution caused by regolith migration on the surface of the asteroid. The topological structures of the equilibria are employed to distinguish the evolutional trends of the local regolith depth, i.e., the local flow map indicates the convergent or emanative behaviors of surface materials, which might explain the formation mechanisms of small-scale ridges, hillocks and pits. We see the prediction of our simplified model agrees with several observed identifications of mass movements based on images returned from the OSIRIS-REx mission, which contributes to our understanding of the long-term trends of regolith movement on Solar System small bodies. Future work will be organized to combine a depth variation model with our global migration model, which is expected to give a better explanation to the small body's reshaping process.

\begin{acknowledgements}
Y.Y. acknowledges financial support provided by the National Natural Science Foundation of China Grant No. 12022212.
\end{acknowledgements}

\bibliographystyle{unsrt}
\bibliography{reference_hcy.bib}   % name your BibTeX data base

\end{document}